\documentclass[3p,times]{elsarticle}

\usepackage{amsmath,amssymb,bm,graphicx}
\usepackage{xcolor}
\usepackage[colorlinks=true,urlcolor=blue,linkcolor=blue,citecolor=blue,bookmarks=false]{hyperref}

\usepackage[figuresright]{rotating}

\begin{document}

\begin{frontmatter}



\title{High-resolution neutron depolarization microscopy of the ferromagnetic transitions in Ni$_3$Al and HgCr$_2$Se$_4$ under pressure}

\author{Pau Jorba}
 \address{Physik-Department, Technische Universit\"{a}t M\"{u}nchen, D-85748 Garching, Germany}

\author{Michael Schulz}
 \address{Heinz-Maier-Leibnitz Zentrum (MLZ), Technische Universit\"{a}t M\"{u}nchen, D-85748 Garching, Germany}

\author{Daniel S. Hussey}
 \address{Physical Measurement Laboratory, National Institute of Standards and Technology, 100 Bureau Dr., MS 8461, Gaithersburg, MD 20899-8461, USA}

\author{Muhammad Abir}
 \address{Nuclear Reactor Laboratory, Massachusetts Institute of Technology, 77 Massachusetts Ave., Cambridge, MA, USA}

\author{Marc Seifert}
 \address{Physik-Department, Technische Universit\"{a}t M\"{u}nchen, D-85748 Garching, Germany}
 \address{Heinz-Maier-Leibnitz Zentrum (MLZ), Technische Universit\"{a}t M\"{u}nchen, D-85748 Garching, Germany}
 
\author{Vladimir Tsurkan} 
 \address{Experimental Physics V, Center for Electronic Correlations and Magnetism, University of Augsburg, D-86159 Augsburg, Germany}
 \address{Institute of Applied Physics, Chisinau, Republic of Moldova}
 
\author{Alois Loidl}
 \address{Experimental Physics V, Center for Electronic Correlations and Magnetism, University of Augsburg, D-86159 Augsburg, Germany}

\author{Christian Pfleiderer}
 \address{Physik-Department, Technische Universit\"{a}t M\"{u}nchen, D-85748 Garching, Germany}

\author{Boris Khaykovich}
 \address{Nuclear Reactor Laboratory, Massachusetts Institute of Technology, 77 Massachusetts Ave., Cambridge, MA, USA}

\begin{abstract}
We performed neutron imaging of ferromagnetic transitions in Ni$_3$Al and HgCr$_2$Se$_4$ crystals. These neutron depolarization measurements revealed bulk magnetic inhomogeneities in the ferromagnetic transition temperature with spatial resolution of about 100~$\mu$m. To obtain such spatial resolution, we employed a novel neutron microscope equipped with Wolter mirrors as a neutron image-forming lens and a focusing neutron guide as a neutron condenser lens. The images of Ni$_3$Al show that the sample does not homogeneously go through the ferromagnetic transition; the improved resolution allowed us to identify a distribution of small grains with slightly off-stoichiometric composition. Additionally, neutron depolarization imaging experiments on the chrome spinel, HgCr$_2$Se$_4$, under pressures up to 15~kbar highlight the advantages of the new technique especially for small samples or sample environments with restricted sample space. The improved spatial resolution enables one to observe domain formation in the sample while decreasing the acquisition time despite having a bulky pressure cell in the beam. 
\end{abstract}


\end{frontmatter}

% \linenumbers




%% References


%%%%%%%%%%%%%%%%%%%%%%%%%%%%%%%%%%%%%%%%%%%%%%%%%

\section{Introduction}

Modern quantum materials exhibit a large variety of interesting phenomena. In these materials, the relative strength of competing interactions can be tuned by non-thermal parameters such as chemical doping or hydrostatic pressure~\cite{si2010heavy,PhysRevLett.93.256404,PhysRevB.86.014422,Yusuf2002}. Close to a quantum critical point (QCP) the balance between the competing interactions allows for strong quantum fluctuations that can stabilize completely new phases, namely non-Fermi liquid behavior~\cite{PhysRevB.69.035111,lohneysen2007fermi,ritz2013formation}, unconventional superconductivity in heavy fermion systems~\cite{si2010heavy,steglich1979superconductivity,mathur1998magnetically}, high-temperature superconductivity in cuprates~\cite{bednorz1986possible,lee2006doping}, and complex magnetic order like in spin ice or spin glasses~\cite{harris1998magnetic,ramirez1994strongly,mirebeau2002pressure}.\\

Two methodical difficulties can arise when a system is driven towards a QCP. First, this fragile new order will be very sensitive to impurities and inhomogeneities, which can mask the intrinsic properties and sometimes even cause new phenomena~\cite{millis2003towards,Dagotto257,pfleiderer2010search}. Traditional bulk thermodynamic measurements, such as magnetometry or specific heat, average over the sample and thus often miss the extra level of complexity. Spatially resolved bulk measurements on the scale of 10~$\mu$m to ~100~$\mu$m are not widely available. Secondly, the sample environment necessary to achieve very low temperatures and high hydrostatic pressures usually lowers the potential to obtain relevant information from the samples, either by making certain techniques completely impossible (e.g. scanning probe measurements) or by strongly decreasing the signal to noise ratio. To alleviate these difficulties, we demonstrate a new method of spatially-resolved thermodynamic measurements of magnetic phase transitions and demonstrate inhomogeneities and domain formation near the transition temperature in Ni$_3$Al and HgCr$_2$Se$_4$, two well-known examples of materials affected by the proximity to a QCP.\\

Ni$_3$Al is an archetypical weak itinerant ferromagnet ~\cite{semadeni2000critical, niklowitz2005spin}. Due to it's proximity to a QCP, the Curie temperature of Ni$_3$Al has strong compositional dependence~\cite{sasakura1984curie}. Well documented inconsistencies in $T_{\mathrm{C}}$ have been attributed to differences in  growth and annealing conditions~\cite{dhar1989low}. Determining the variation of $T_{\mathrm{C}}$ across a specimen is important, since the composition may vary due to diffusive processes during growth. As a consequence, bulk magnetometry measurements or neutron scattering experiments which effectively average over the entire sample volume may give inaccurate results. 

HgCr$_2$Se$_4$ belongs to the isostructural family of the chromium spinels ($A\mathrm{Cr}_2X_4$). It shows diverse magnetic ground states mostly due to the competition of the direct antiferromagnetic exchange between the chromium ions, and the ferromagnetic superexchange mediated by the $X$ nonmagnetic atoms~\cite{baltzer1966exchange, yaresko2008electronic, rudolf2007spin}. The ferromagnetic superexchange dominates at large inter-ionic distances, while it gives way to the antiferromagnetic exchange at smaller distances.~\cite{baltzer1966exchange}. Hydrostatic pressure is a unique tuning parameter to study the mechanisms of these competing interactions. In HgCr$_2$Se$_4$ ferromagnetism is suppressed under pressure at a rate of 0.95~K/kbar from an initial transition temperature of 105~K~\cite{srivastava1969pressure}. The influence of stoichiometry and pressure inhomogeneities on the transition temperature are not well known and difficult to study with standard techniques. 

Neutron depolarization imaging (NDI) provides a spatially-resolved thermodynamic probe of a sample's magnetic state~\cite{mschulz_ICNS2009, schulz2016neutron}, and potentially a three-dimensional distribution of the magnetic field~\cite{rekveldt1973study, kardjilov2008three, schulz2010towards, treimer2014radiography}.  Neutrons readily penetrate cryogenic equipment, pressure cells and bulk metal samples. When a neutron non-adiabatically enters a magnetic field, its magnetic moment undergoes Larmor precession. Changes in the polarization state of the transmitted neutron beam can probe the magnetic domain distribution inside a ferromagnet~\cite{halpern1941passage, rekveldt1973study, schlenker1973polarized}, magnetic islands in spin glasses~\cite{mirebeau1990neutron, mitsuda1992neutron, Yusuf2002}, or the Meissner field outside a superconductor~\cite{weber1974properties, roest1993three, treimer2013imaging}. 

Conventional NDI is based on pinhole optics where a collimated beam produced by an aperture of characteristic size $D$ ($\sim$1~cm), illuminates a sample placed a distance $z$ from the detector, with overall length $L$ ($\sim$10~m) from the aperture. The geometric image blur is approximately $\lambda_{\mathrm{g}} \approx D z / L$, and is significant in NDI. To achieve magnetic contrast, a neutron polarizer and spin flipper are placed upstream of the sample, while a neutron spin analyzer is placed between the sample and the detector. To accommodate the spin analyzer between the sample and detector requires $z \sim 0.5$~m. To obtain a quantitative measure of the sample's magnetic state, the neutron beam must be monochromatized, reducing the intensity and requiring larger pinhole diameters to achieve practical image acquisition times. As a result the typical blur $\lambda_{\mathrm{g}} \sim$~1~mm.  In contrast, samples as small as 1~mm$^3$ are often necessary either because larger crystals are not available or sample space limitations of sample environment equipment (e.g. pressure cells) exist so that only bulk-average measurements are possible.

To overcome these shortcomings, this study presents the first polarized neutron microscope. The measurements were performed at the ANTARES instrument at Heinz Maier-Leibnitz Zentrum (MLZ)~\cite{calzada2009new, antares_JLSRF}. Focusing neutron mirrors in a Wolter optic configuration with focal lengths in the meter range~\cite{khaykovich2011x, liu2013microscope} enabled high resolution images in combination with the required optical components for NDI. The unprecedented spatial resolution, of about 100~$\mathrm{\mu}$m, is determined by the angular resolution of the mirrors. 

\section{Experimental details}
\subsection{Instrument Configuration}
\begin{figure}
\includegraphics[width=1.0\linewidth]{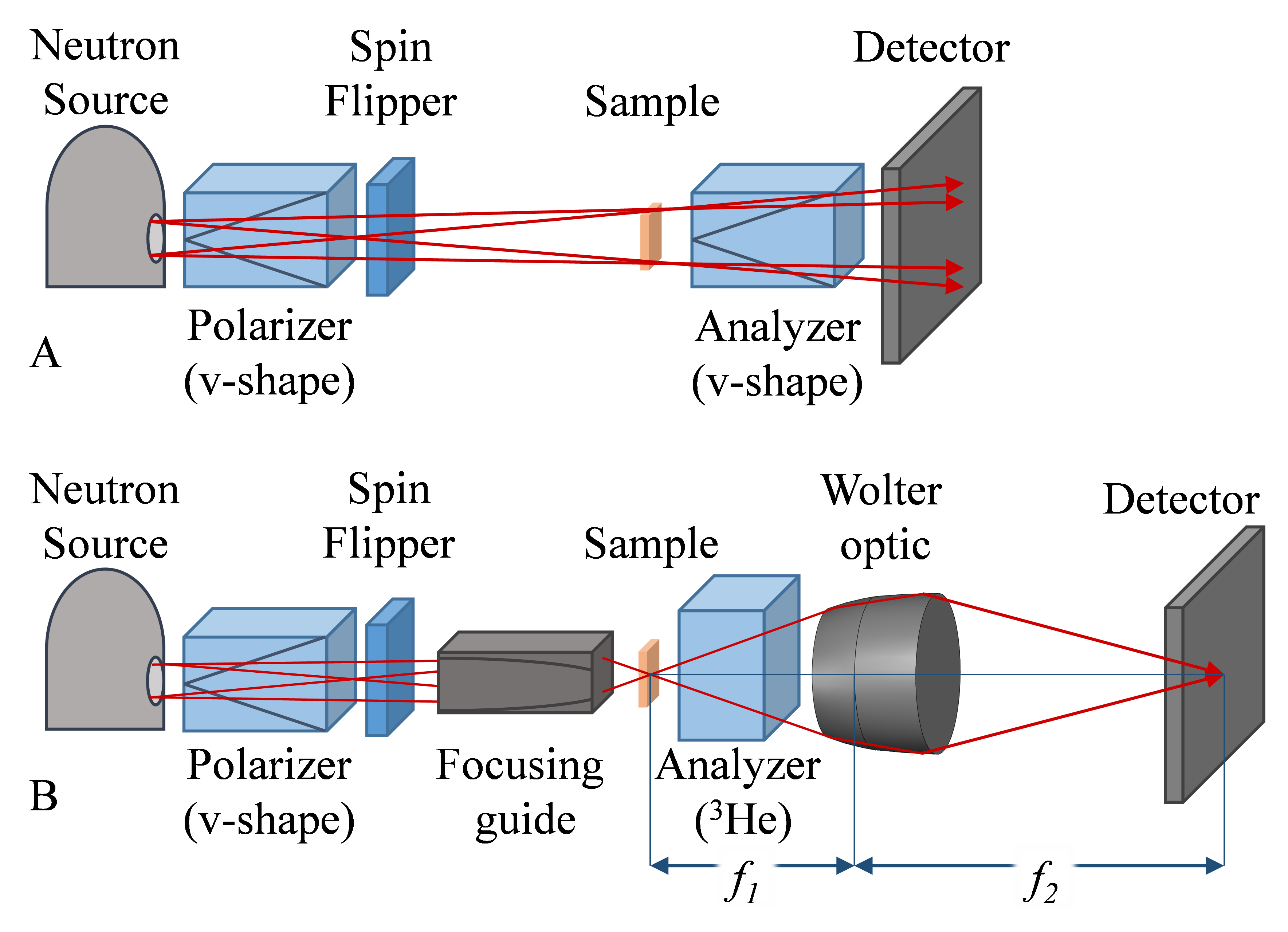}
\caption{Schematic of the neutron imaging techniques used in this study. (A)~Standard neutron depolarization imaging. Red arrows illustrate neutron flight paths highlighting the geometrical unsharpeness due to the separation of sample and detector. (B)~Polarized neutron microscope. A focusing guide increases the divergence of the neutron beam while a Wolter optic with focal lengths $f_{1}$ and $f_{2}$ behind the sample acts as a magnifying lens. A $^3\mathrm{He}$ analyzer is placed between the sample and the optic without effecting the spatial resolution.}
\label{figure1}
\end{figure}

Layout schematics of a standard NDI and the microscope are shown in Fig.~\ref{figure1}. The source is a variable-diameter beam aperture (36~mm, 18~mm, 9~mm) at the entrance to the experimental hutch. The sample is mounted in a Sumitomo SRDK-205D closed-cycle refrigerator (CCR)~\cite{NIST}. 
CCR's are standard equipment used to study temperature-induced transitions. The temperature homogeneity across the sample was assured by clamping the sample on an Aluminum support which has proven to provide excellent temperature stability in prior experiments.

In standard NDI, the source is projected onto the detector, situated 9~m from the aperture, modulated by the sample. 
For the microscope, a focusing guide was placed 8~m from the entrance aperture (Fig.~\ref{figure1}B). The Wolter optic accepts a large divergence, so the largest available aperture (36 mm) was employed without the loss of spatial resolution. The sample was located 50~mm downstream from the focusing guide, close to its focal spot. The relative positions of the Wolter optic and the detector were adjusted to have the sample at the object plane $f_1$ of the optic and the detector at the focal plane $f_2$. This arrangement corresponds to a basic microscope; the focusing guide is a condensing lens; the Wolter optic is the magnifying, image-forming lens. 

Wolter optics have been previously described and demonstrated to form neutron images~\cite{khaykovich2011x, liu2013microscope}. The optic employed here has a magnification of 4, total focal length of 3.2~m, $f_1$ = 0.64~m, a diameter of about 3.5~cm, and overall length of about 6~cm~\cite{liu2013microscope, liu2013sans}.  The three 0.2~mm thick foil mirrors are composed of pure nickel, so that neutrons of 5~{\AA} incident on the mirror surface at an angle less than about $0.5^\circ$ are totally reflected.  The angle of the first mirror is about $1^\circ$, so that the divergence acceptance of the optics is 1$^\circ$ to 1.5$^\circ$.   

The focusing neutron guide, made by SwissNeutronics~\cite{boni2008new, komarek2011parabolic, adams2014versatile}, consists of four parabolic sections with a nickel/titanium supermirror coating with $m = 6$ in a geometry so that the initially collimated imaging beam was focused to a spot size of about 1~mm radius.
The focal distance is 80~mm from the exit of the guide, 
creating a beam divergence of 0.75$^\circ$ to 2$^\circ$ that reasonably fills the acceptance of the Wolter optic. 
The separation of the focusing guide and the Wolter optic was adjusted to achieve the largest uniform field of view which was about $2\times2~\mathrm{mm}^2$. 
%Michi:
It should be noted that the Wolter optic could be used with a larger beam cross section, which could however not be achieved with the available focusing guide.

\subsection{Field of view}

\begin{figure}
\includegraphics[width=1.0\linewidth]{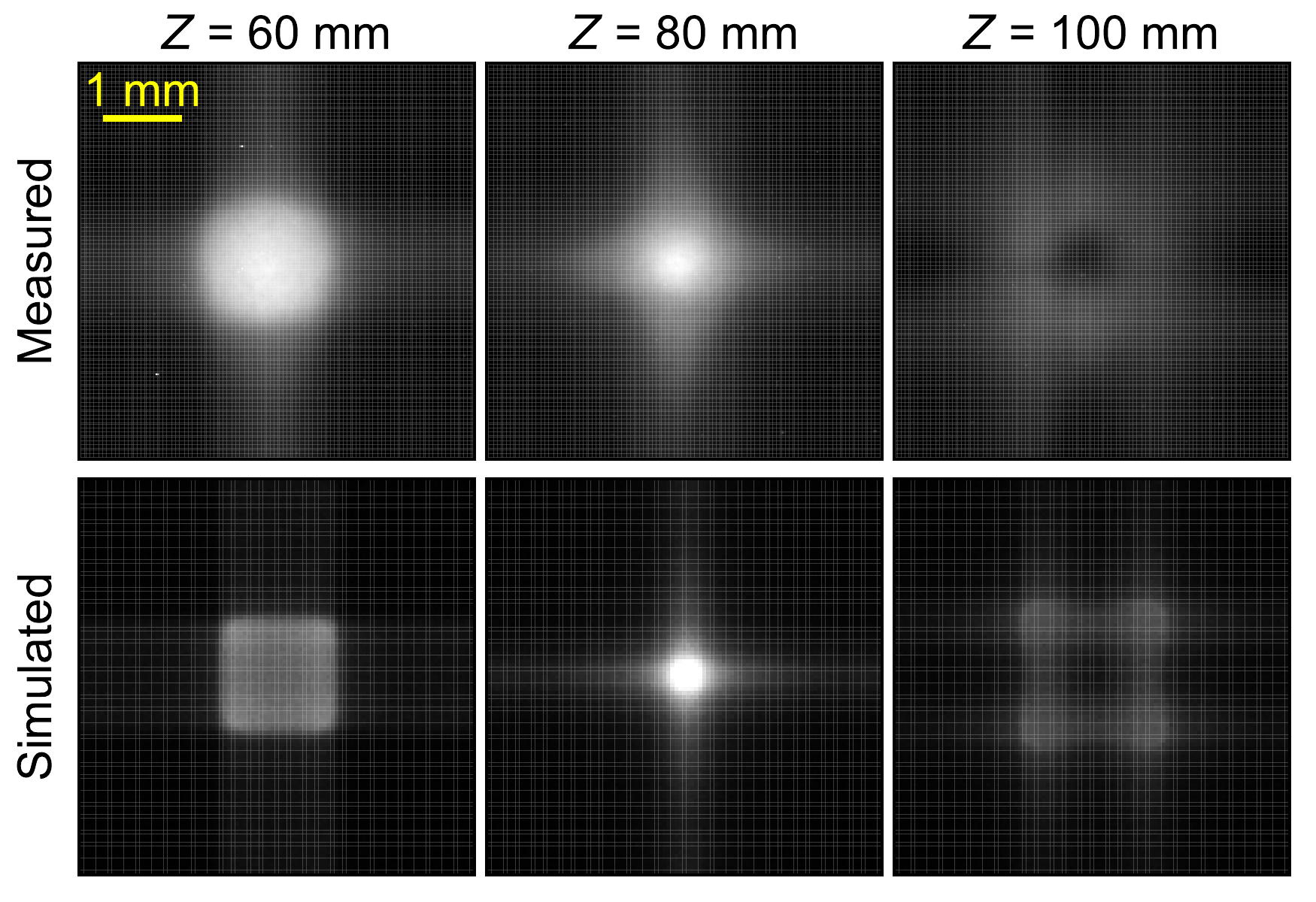}
\caption{Measured (top row) and simulated (bottom row) neutron beam intensity at 3 different positions \textit{Z} downstream of the focusing guide. Higher beam intensities are denoted by lighter areas. The grey scale of the measured and the simulated images has been normalized based on the highest observed neutron count. The spatial scale is the same for all images.}
\label{figureS1}
\end{figure}

In order to generate a neutron image with the Wolter optic, the sample must be illuminated by a divergent beam. Fig.~\ref{figureS1} shows a comparison of the measured and simulated neutron intensity distribution at different positions downstream from the focusing guide. The measured radiographs (top row) were obtained with a simplified version of the setup shown in Fig.~1B, where the polarizer, spin flipper, analyzer, and sample were removed. The focusing guide was translated along the beam direction while the Wolter optic was maintained at a fixed distance ($f_2$) from the detector, thus generating images of the beam section at different distances \textit{Z} from the exit of the focusing guide. The simulations (bottom row) were obtained using McStas~\cite{lefmann1999mcstas, willendrup2004mcstas}, a neutron-ray tracing package.  The simulation was performed by adapting a library that simulates the ANTARES instrument into the configuration of the current experiment with a collimator at the entrance of the instrument (with 36~mm diameter), a velocity selector tuned to 5~{\AA}, and a parabolic focusing guide at 8~m from the collimator. The focusing guide in the ray tracing model was defined to match the m = 6 coating, dimensions, and focal distance of the one used in the experiment. The images were generated from 10$^9$ virtual neutrons with a 2D virtual detector placed at different distances \textit{Z} from the exit of the focusing guide. 

The measured and simulated images show good agreement. Note that the simulations only included the focusing guide and stopped at the sample position, while in the experiment, the image of the beam intensity at the sample position was formed by the Wolter mirrors. The unsharpness and distortion of the measured images as compared to the simulated ones is due to the finite focal depth of the Wolter optics, and optical aberrations~\cite{khaykovich2011x,liu2013microscope}. Nonetheless, the simulations are in excellent agreement with the measurements. When the neutron beam exits the focusing guide it shows a homogeneous distribution with decreasing square shaped cross section until the focal length of the focusing guides ($Z = 80~\mathrm{mm}$) where the beam starts diverging again. Downstream of the focal spot of the focusing guide ($Z > 80~\mathrm{mm}$), the beam distribution is no longer homogeneous.  The focusing guide increases the divergence of the neutron beam to values ranging from $0.75^\circ$ to $2^\circ$, which matches very well the divergence acceptance of the Wolter optics. It is important here to remark that the use of the focusing optics not only increases the beam divergence, but also increases significantly the neutron flux at the sample position, thus decreasing the exposure time. As a result of these simulations, in all the measurements using the polarized neutron microscope, the samples were placed about 50~mm downstream from the exit of the focusing guide in order to produce the largest and most uniform region of illumination which was about $2\times2~\mathrm{mm}^2$.

\begin{figure}
\includegraphics[width=1.0\linewidth]{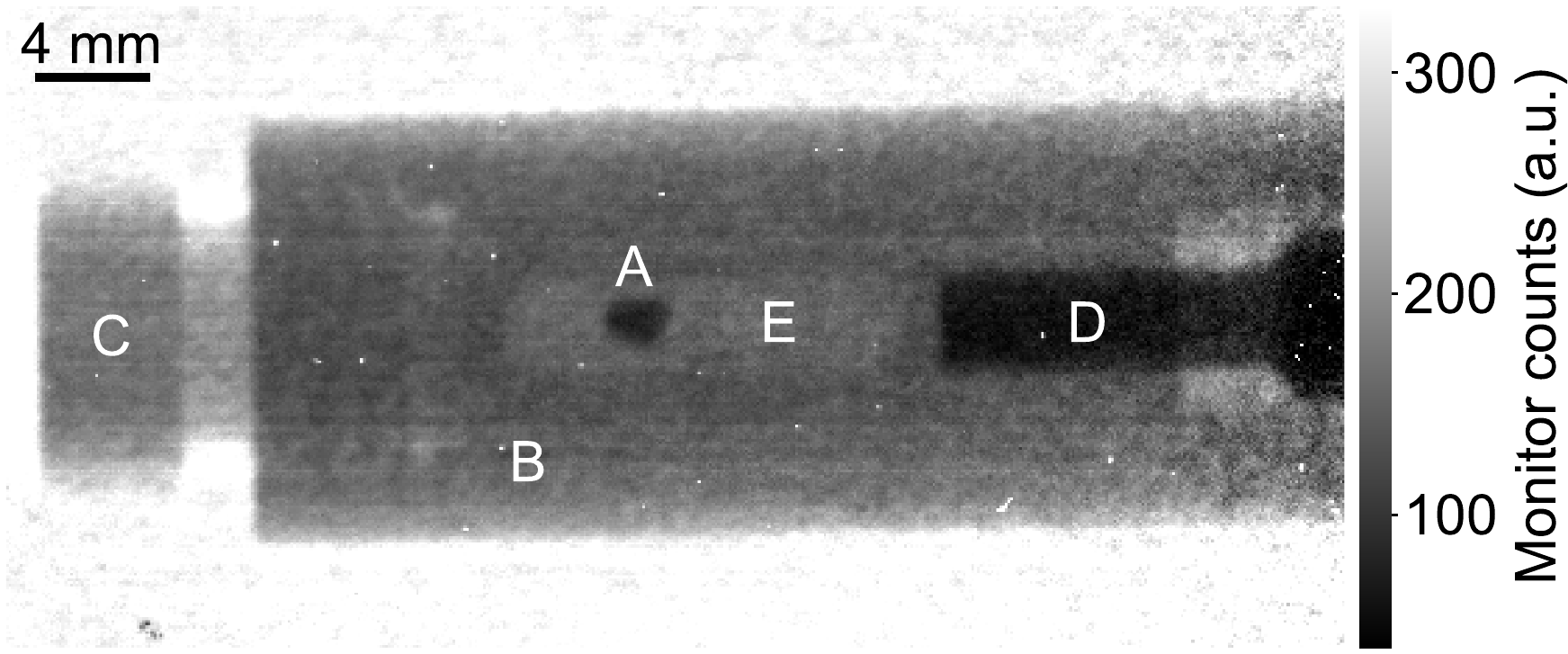}
\caption{Conventional, flat-field normalized neutron radiography of the clamp cell used to pressurize the HgCr$_2$Se$_4$ sample (A). The cell body (B) and the locking nut (C) are made of hardened copper-beryllium. A tungsten-carbide piston (D) is pressed into the sample space (E) filled with a liquid pressure medium.}
\label{figureS2}
\end{figure}

\subsection{Neutron beamline equipment}
A neutron velocity selector~\cite{friedrich1989high} defined a wavelength band with $\Delta\lambda/\lambda \approx$~10~$\%$.  The detector consisted of an Andor iKON-L CCD, which viewed a 200~$\mu$m 
thick LiF:ZnS scintillator screen via a Zeiss 100~mm lens with f~1.8, yielding an effective pixel pitch of about 78~$\mu$m. 
The beam was polarized by a transmission v-cavity polarizer that was placed before the focusing guide. A Mezei type spin flipper was placed 0.5 m downstream of the polarizer 
allowing a $\pi$ spin rotation. A vertical guide field of about 0.5~mT ensured there were no zero-field crossings between the polarizer, flipper and focusing guide. After passing through the sample, the neutrons were spin-analyzed by a $^3$He gas which was polarized at the HELIOS facility at MLZ~\cite{hutanu2007scientific}. $^3$He was employed as a spin analyzer so as to not introduce additional structures in the image of the divergent beam.  The polarization, $P_{\mathrm{He}}$, of the $^3$He gas in the analyzer was measured while the sample temperature was changed by acquiring flipping ratio measurements without the sample in the field of view and used to correct the polarization images of the sample. At the beginning of the measurement, the flipping ratio was about 5.89 decaying to 3.5 overnight; the $^3$He gas cell was replaced about once every 24 hours to maintain a reasonable flipping ratio. Polarization images were calculated by combining two images with the same exposure time, with the spin flipper turned on ($I_{down}$) or off ($I_{up}$): $P=(I_{up}-I_{down})/(I_{up}+I_{down}-2I_{dark})$. $I_{dark}$ is an offset image taken with the beam shutter closed. To minimize non-statistical sources of noise, each of the image sets are obtained by computing the median of three images.

\subsection{Clamp Cell}
The clamp cell body is a hollow cylinder of copper-beryllium  with 12~mm outer diameter and 3~mm inner diameter. A conventional radiograph is shown in Fig.~\ref{figureS2}. A locking nut is screwed in one of the ends of the cell body, while a tungsten carbide piston pushes into the sample space. A Teflon capsule is tightly fit in the sample space, enclosing the sample and the pressure medium, a 1:1 mixture of Fluorinert FC 72 and FC 84. Fluorinert is a fluorocarbon-based fluid and is very transparent to neutrons. The neutron transmission through the cell body and sample space filled with pressure medium is about 46~\%. The sample has a stronger contrast due to the large neutron absorption cross section of mercury.

\section{Experimental results and discussions}

\subsection{Ni$_3$Al}
\begin{figure}
\includegraphics[width=1.0\linewidth]{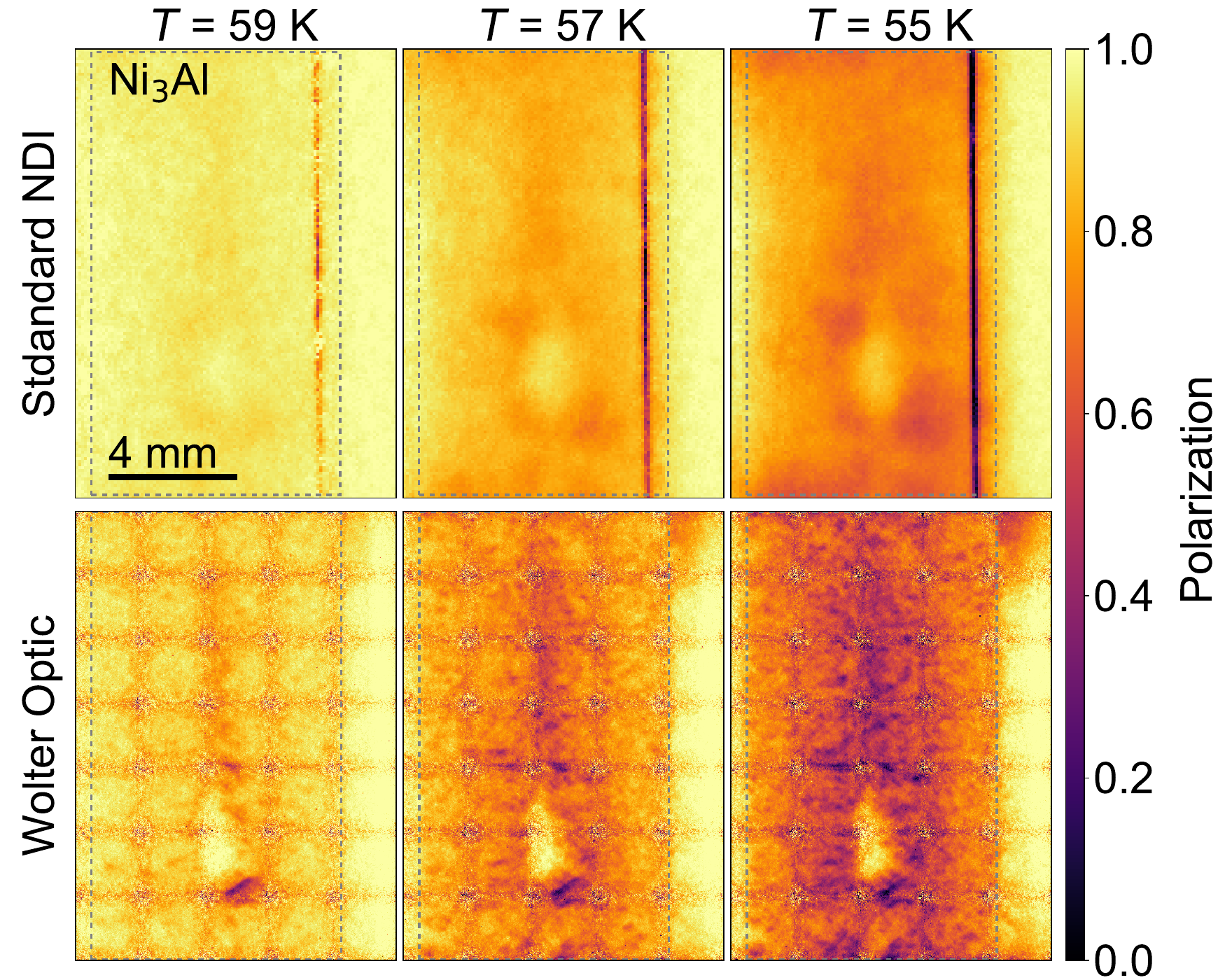}
\caption{Neutron polarization images of a Ni$_3$Al sample at 59 K, 57 K and 55 K. The top row are images obtained with standard neutron depolarization imaging. The bottom row images are the result of stitching 35 images obtained with the polarized neutron microscope. The dashed square delimits the sample edges. }
\label{figure2}
\end{figure}

A rectangular slab ($8\times14\times2~\mathrm{mm}^3$) of Ni$_3$Al was obtained from a stoichiometric Ni$_3$Al-rod cast from high-purity starting elements using an inductively heated furnace. The rod was annealed at $973~^{\circ}\mathrm{C}$ for 150~h and subsequently at $773~^{\circ}\mathrm{C}$ for 8.3~h in a Knudsen-type effusion cell in a 1.1~bar argon atmosphere with less than 0.001 ppm impurities. The sample was annealed in the hope that local concentration variations would diminish and the overall homogeneity would improve. 
The sample was measured at three temperatures around $T_\mathrm{C}$. In Fig.~\ref{figure2} we compare images obtained from standard NDI (top row) with those obtained with the Wolter optic (bottom row). The sample outline is indicated by a dashed line. The standard NDI experiment used a pinhole of 18 mm diameter ($L/D = 470$), and the velocity selector was set to 4.3~{\AA}, enabling a polarization image acquisition time of 180~s. For the microscope measurements the velocity selector was tuned to 5~{\AA} with with 360~s acquisition time for a single polarized neutron image. Since the largest nearly uniform intensity pattern of the focusing guide was about $2\times2~\mathrm{mm}^2$, the sample was rastered by moving the cryostat containing the sample in 2 mm steps in both directions perpendicular to the beam to image the entire sample. This procedure required 35 frames and 3~h 30~min per temperature. 

The v-shaped polarizers used in the standard NDI experiment suffer from poor polarization at the supermirror junctions, leading to a vertical line artifact in the right hand side of the images. The images obtained with the Wolter optic show a clear stitching artifact  due to the lower counting statistics at the edge of the field of view of a single frame. It would be possible to alleviate this artifact by employing a larger area of uniform intensity at the sample position or taking smaller raster steps.
The neutron microscope images replicate the general features of the standard NDI experiment. In particular  an oval shaped, weakly magnetic region with higher polarization is observed in the lower third of the sample.
The heterogeneities are better resolved with the microscope, revealing many smaller grains having high depolarization values. 
This is evident even in the presence of the stitching artifact.
The polarization values are on average a bit higher for the standard NDI experiments as neutrons with shorter wavelengths are less depolarized for a given magnetic field~\cite{halpern1941passage}.  While the collimation values used for the NDI experiment yield $\lambda_{\mathrm{g}}$~=~1~mm, the polarized neutron microscope provides  a factor 10 improvement in spatial resolution.

From the Al-Ni binary alloy phase diagram~\cite{okamoto2004ni} it is evident that crystallizing a stoichiometrically pure Ni$_3$Al sample from a melt poses challenges. During cooling many small composition inhomogeneities will appear. Small changes in the stoichiometry have strong effects on the magnetic properties. There is a sharp decrease of the critical temperature and ordered magnetic moment with excess aluminum, and a strong increase with nickel excess~\cite{sasakura1984curie, idzikowski2006magnetic}. The magnetic moment of the nickel atoms in the stoichiometrically pure Ni$_3$Al samples is ${\sim}0.3$~$\mu_{\mathrm{B}}$ per atom; a change of $\pm$1\% in nickel concentration changes the magnetic moment by $\pm$0.1~$\mu_{\mathrm{B}}$~\cite{velasco2015short}. The beam depolarization depends exponentially on the magnetic field inside a magnetic domain~\cite{halpern1941passage}. Thus, we can then identify the weakly depolarized (P~$\ge$~0.8) areas in Fig.~\ref{figure2} with aluminum rich regions, and strongly depolarized areas (P~$\le$~0.2) as nickel rich regions. This information can be employed to adjust the synthesis parameters of these alloys. For instance, annealing is expected to reduce inhomogeneity in the composition and narrow down short range order fluctuations, increasing the ordered magnetic moment while decreasing the critical temperature~\cite{van1977structural,velasco2015short}. By observing samples with different annealing histories we could get a unique insight into the diffusion processes taking place in this alloy. Moreover, by performing measurements with finer temperature steps the distribution of $T_{\mathrm{C}}$ and the Ni concentration across the sample could be determined. With this information smaller samples with defined magnetic properties could be prepared from the large sample and used e.g. for bulk magnetization measurements.

\subsection{HgCr$_2$Se$_4$}

\begin{figure}
\includegraphics[width=\linewidth]{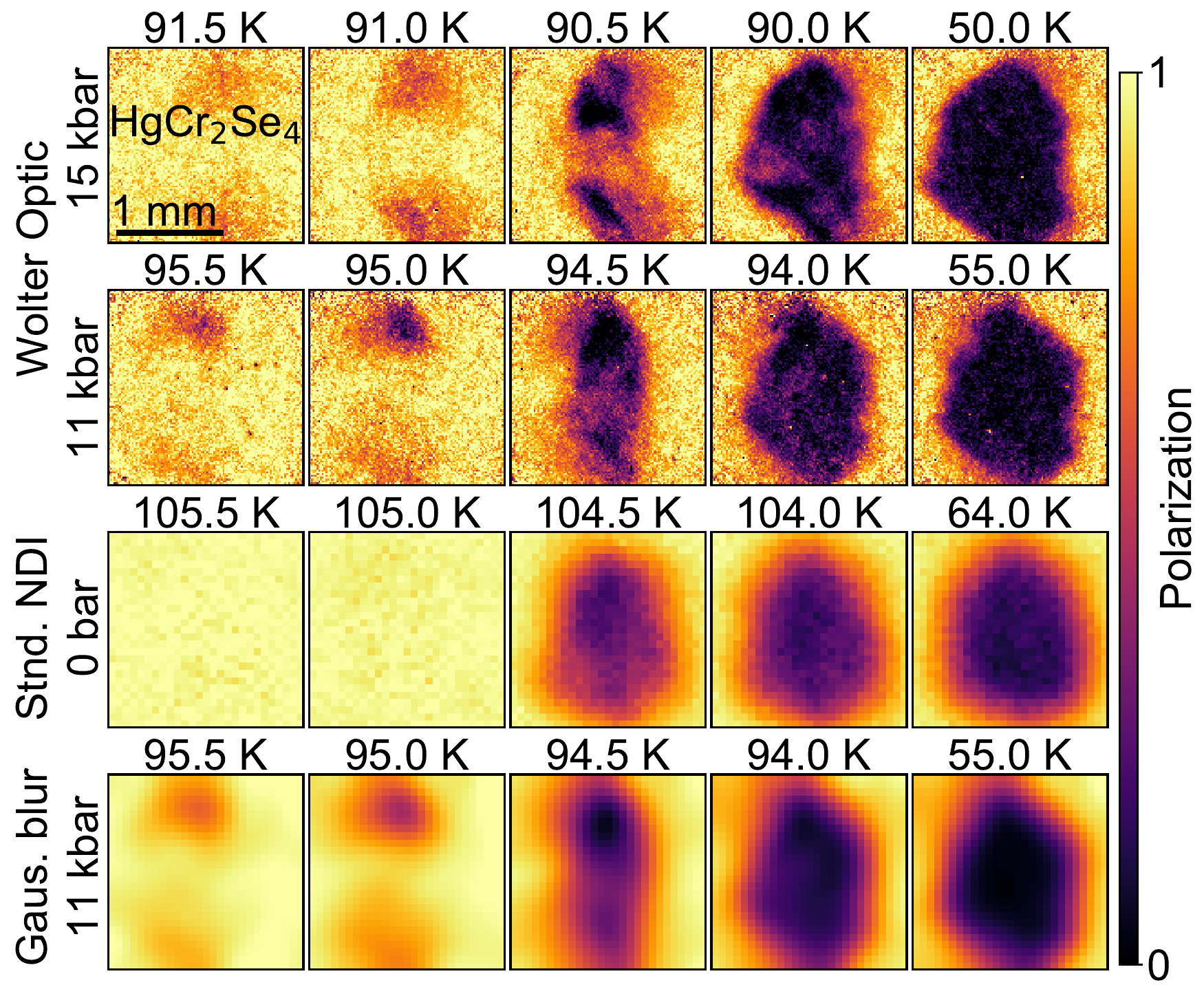}
\caption{Neutron polarization images of a HgCr$_2$Se$_4$ sample at four temperatures around the critical temperature with 0.5 K steps (first 4 columns), and deep in the ferromagnetic phase (rightmost column). The images in the first 2 rows were obtained with the polarized neutron microscope under hydrostatic pressure. The first and second row correspond to images taken at 15~kbar and 11~kbar of pressure respectively. The images in the third row were obtained at ambient pressure via standard NDI. In the bottom row, we show the same data as in the experiment at 11 kbar (2nd row) where we have applied a Gaussian filter with $2\sigma = 0.5~\mathrm{mm}$ and rebinned the image to match the standard NDI resolution. Color and spatial scales are the same for all the images. Note that the sample orientation differs between the applied pressures.}
\label{figure3}
\end{figure}

The HgCr$_2$Se$_4$ sample was an irregularly shaped single crystal of ${\sim}1~\mathrm{mm}^{3}$ volume, grown by means of chemical transport  from pre-synthesized high-purity polycrystals, and was pressurized in a copper-beryllium clamp cell, described above. In Fig.~\ref{figure3} we summarize the polarization images of the crystal under 11~kbar and 15~kbar of hydrostatic pressure with the microscope (first and second row), and at ambient pressure using conventional NDI (third row). The sample fit in the field of view produced by the focusing guide; without the need for rastering, the acquisition time was reduced from a few hours to a few minutes per temperature. The temperature was scanned in finer steps around $T_{\mathrm{C}}$, and through the whole range of the ferromagnetic phase transition. No hysteresis in $T_{\mathrm{C}}$ was observed, confirming that the sample was properly thermalized. For the standard NDI experiment a beam aperture of 9 mm was used, giving a collimation value of $L/D=944$ and the acquisition time for an image was 600~s. With the polarized neutron microscope, the acquisition time was 360~s at 11~kbar and 600~s at 15~kbar. The sample orientation was not carefully maintained when the pressure was changed and differs across the different series of images.

\begin{figure}
\includegraphics[width=\linewidth]{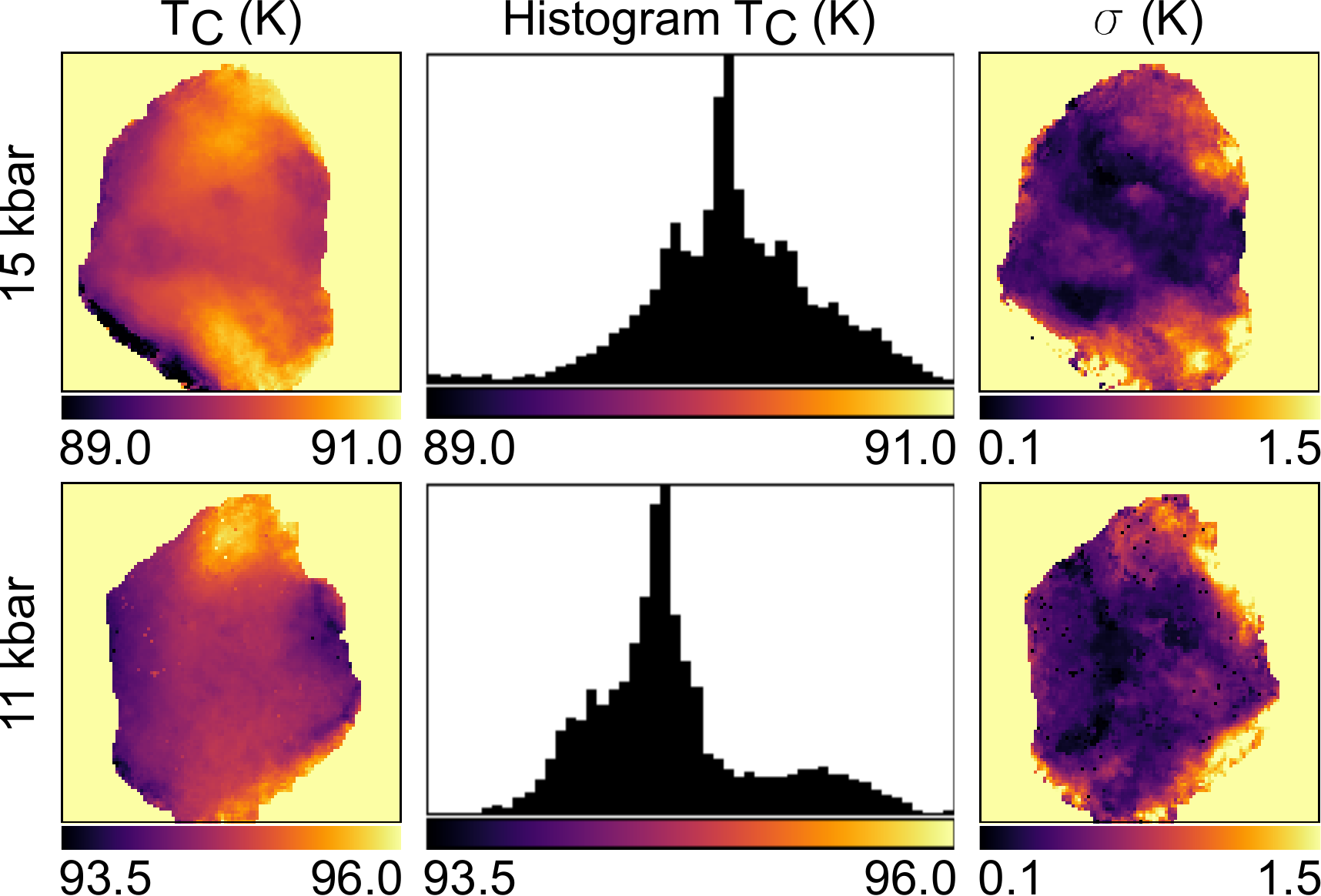}
\caption{Distributions of $T_\mathrm{C}$ in the HgCr$_2$Se$_4$ sample for 11 kbar and 15 kbar. Spatial distribution is in the first (left) column. Histograms of $T_\mathrm{C}$ over the sample are in the second column. The histograms are used to determine the value of $T_\mathrm{C}$. The third column shows the fit parameter $\sigma$ of the error function, which can be interpreted as the variation of $T_\mathrm{C}$ along the beam direction for each pixel. The values of $\sigma$ are much larger near the edges than in the bulk, as expected due to defects and inhomogeneities.}
\label{fig_Tc_hor}
\end{figure}

As a first observation, the increase of spatial resolution from the standard NDI to the polarized neutron microscope images is significant. 
From the images deep in the ferromagnetic phase (Fig.~\ref{figure3}, right column) the sample outline is much sharper with the neutron microscope. The calculated resolution of the standard NDI experiment is $\lambda_g \approx 500~\mathrm{\mu m}$, the highest resolution routinely achieved in NDI~\cite{mschulz_ICNS2009, Kardjilov2009nuclear}.  While standard NDI severely suffers from the flux loss resulting from smaller pinholes used in order to increase the spatial resolution, the polarized neutron microscope yields a 5 fold increase in spatial resolution to $100~\mathrm{\mu m}$, while at the same time nearly halving the exposure time for this experiment even despite the additional absorption by the pressure cell. The image quality does not suffer from the lower counting time at 11~kbar. For comparison of spatial resolution the last row of Fig.~\ref{figure3} shows the neutron microscope data at 11~kbar after smearing with a Gaussian filter with $2\sigma = 0.5~\mathrm{mm}$ and rebinning the images to match the standard NDI resolution. It is clearly observed that the blurred microscope data and the standard NDI data show similar spatial resolution despite the different sample orientation in the measurements. Many of the smaller features e.g. at the edges of the sample are lost after smearing the microscope data which obviously shows the benefit of increased spatial resolution.

In order to accurately determine the Curie temperature of the sample and compare the values at different pressures we fitted the polarization data for each pixel with an error function
\[A+\frac{B}{2}\left( 1 + erf\left(\frac{x-T_\mathrm{C}}{\sqrt{2}\sigma}\right) \right),\]
where A is the offset, B the amplitude of the error function and $\sigma$ is the standard deviation which can be interpreted as the variation in $T_\mathrm{C}$. Fig.~\ref{fig_Tc_hor} shows the obtained $T_C$ images for 11 kbar and 15 kbar.

The change of $T_{\mathrm{C}}$ from 105~K at ambient pressure to 94.5~K and 90.5~K at 11~kbar and 15~kbar, respectively, confirms that the linear decrease of $T_\mathrm{C}$ with pressure reported in Ref.~\cite{srivastava1969pressure} is still valid up to 15 kbar. The histograms in Fig.~\ref{fig_Tc_hor} show that the distribution of $T_\mathrm{C}$ is shifted in pressure with only minor change in the shape of the distribution. At ambient pressure the critical temperature is uniform, but for the pressure experiments the polarization starts to drop 1~K above $T_{\mathrm{C}}$ at the top and bottom of the sample as observed in the polarization images in Fig.~\ref{figure3} and also confirmed by the fitted $T_\mathrm{C}$ values in Fig.~\ref{fig_Tc_hor}. When the temperature is well below $T_{\mathrm{C}}$, we observe that the depolarization extends uniformly through the sample. The main reasons for this distribution in $T_{\mathrm{C}}$ could be inhomogeneities in temperature, composition or pressure. The distribution of $T_C$ values along beam direction as determined by the values of $\sigma$ in Fig.~\ref{fig_Tc_hor} is small in the center of the sample and more pronounced at the edges. 
We can exclude an inhomogeneous temperature distribution over the sample since a homogeneous gradient across the sample would be expected from the fact that the pressure cell is connected to the cold finger only at the top. As a consequence also the observed $T_{\mathrm{C}}$ should show a gradient from top to bottom, which is not observed.
In contrast to Ni$_3$Al, the effect of compositional impurities on $T_{\mathrm{C}}$ is small as demonstrated by the small spread of values reported in the literature~\cite{baltzer1966exchange, rudolf2007spin, srivastava1969pressure}. Furthermore, the fact that at ambient pressure no such spread in $T_{\mathrm{C}}$ is observed, leads us to reject a compositional distribution. The employed pressure medium (Fluorinert) freezes at 130~K at ambient pressure, and the freezing point increases linearly with pressure, reaching room temperature at ${\sim}18$ kbar~\cite{torikachvili2015solidification}. The uneven freezing of the pressure medium upon cooling of the pressure cell can lead to anisotropy and unequal distribution of pressures around a mean value. We therefore assume that a pressure distribution is the most likely cause for the inhomogeneous $T_{\mathrm{C}}$ distribution. 

In the images in Fig.~\ref{figure3} at 11 kbar and 15 kbar, just below the critical temperature (94~K and 90~K respectively) a patched pattern is observed. This could be evidence of domain formation. At $T_{\mathrm{C}}$, as the ferromagnet softens, the guide field of about 5~mT is enough to saturate the magnetization. Deep in the ferromagnetic phase, the mean domain size is much smaller than the spatial resolution of the techniques described here (typically 1-10~$\mathrm{\mu}$m) and 
the strong magnetic moment of the chromium atoms ($\approx 6~\mathrm{\mu}_{\mathrm{B}}$) in combination with the large sample thickness lead to an almost fully depolarized beam  (Fig.~\ref{figure3} right column). Between the single domain situation at the critical temperature, and the microscopically small domains at lower temperatures, there must be an intermediate state where a few macroscopic domains are distributed across the sample. The observed pattern could correspond to the distribution of a few magnetic domains with sizes of around 200~$\mu$m. Again, due to the poor spatial resolution of standard NDI, we are unable to make the same assessment for the image at ambient pressure and 104~K. 

\section{Conclusions and outlook}

In conclusion, we have demonstrated a polarized neutron microscope with a spatial resolution of about 100~$\mu$m which is up to a factor 10 improvement over conventional NDI, by combining condensing and image-forming optics (focusing guide and Wolter mirrors). Due to the large focal length of the Wolter optic, high spatial resolution is achieved even with the use of a cryostat and polarized $^3$He neutron spin analyzer between the sample and detector. Moreover, Wolter optics can magnify the neutron image to further improve the resolution beyond the intrinsic capabilities of the detector and take the NDI technique to an unprecedented level of spatial resolution.

The improved spatial resolution of the depolarization images of a Ni$_3$Al sample allows us to identify stoichiometrically inhomogenous regions clustered into grains of a few hundred micrometers. This could enable us to optimize the preparation process of such materials or even pick parts prepared from the original specimen with enhanced homogeneity for bulk magnetization measurements.

Detailed polarization images were obtained for a HgCr$_2$Se$_4$ sample inside a beryllium-copper clamp cell  at 11~kbar and 15~kbar. In addition, we mapped both the distribution of $T_\mathrm{C}$ over the sample and the homogeneity along the beam direction. Thus, effectively three-dimensional information about the magnetic properties of the sample can be obtained with relatively short experimental time. The linear decrease of the critical temperature under applied hydrostatic pressure was confirmed up to 15~kbar. With the aid of the Wolter optic, features of ${\sim}100~\mathrm{\mu}$m were resolved, which was hitherto not achievable with standard NDI. At the same time, it was possible to decrease the acquisition time to almost the half despite the additional beam absorption due to the pressure cell.

The mirrors used in this experiment were made for the ease of demonstration and were not optimized for the ANTARES beamline. Already this prototype device clearly shows the advantage of a microscope over pinhole optics for NDI. We expect that an optimized Wolter optic would lead to unprecedented capabilities in polarized imaging such as the observation of ferromagnetic domains in the bulk of a sample. In particular, anticipated advances in mirror fabrication would enable obtaining spatial resolution of about 10~$\mathrm{\mu}$m over a 1 cm field of view, resulting in a factor 100 improvement over conventional NDI.  As well, it would be possible to nest more than three mirror shells yielding even larger gains in time resolution.

In order to better understand the complexities of quantum materials, often spanning length scales from nanoscopic over microscopic to mesoscopic, technical developments as the one here presented can prove extremely valuable.

%\end{document}

\section{Acknowledgments}

The authors are grateful to S. Masalovich for providing the polarized $^3$He gas, and to Philipp Schmakat for creating the McStas library simulating the ANTARES beamline on which the simulations of the beam profile have been based. We thank P. B{\"o}ni for providing and advising on the use of the parabolic focusing supermirror guides. The work at MIT was performed under the following financial assistance award 60NANB15D361 from U.S. Department of Commerce, National Institute of Standards and Technology. Work partly supported by the DFG via the Transregional Collaborative Research Center TRR 80. This work is based upon experiments performed at the ANTARES instrument operated by FRM II at the Heinz Maier-Leibnitz Zentrum (MLZ), Garching, Germany.

\bibliographystyle{elsarticle-num}
\bibliography{wolter}

\begin{thebibliography}{10}
\bibitem{NIST}
Certain trade names and company products are mentioned in the text or
  identified in an illustration in order to adequately specify the experimental
  procedure and equipment used. In no case does such identification imply
  recommendation or endorsement by the National Institute of Standards and
  Technology, nor does it imply that the products are necessarily the best
  available for the purpose.
\end{thebibliography}

\end{document}